# Improving Network Efficiency by Removing Energy Holes in WSNs

M. B. Rasheedl[1], N. Javaid[1], A. Javaid[2], M. A. Khan[1], S. H. Bouk[1], Z. A. Khan[3]

[1]COMSATS Institute of Information Technology, Islamabad, Pakistan.
[2]Mathematics Dept., COMSATS Institute of Information Technology, Wah Cant., Pakistan.
[3]Faculty of Engineering, Dalhousie University, Halifax, Canada.
Email: nadeemjavaid@comsats.edu.pk, Phone: +92519049323

Abstract— Cluster based Wireless Sensor Networks (WSNs) have been widely used for better performance in terms of energy efficiency. Efficient use of energy is challenging task of designing these protocols. Energy holedare created due to quickly drain the energy of a few nodes due to non-uniform distribution in the network. Normally, energy holes make the data routing failure when nodes transmit data back to the base station. We proposedEnergy-efficient HOleRemoving Mechanism (E-HORM) technique to remove energy holes. In this technique, we use sleep and awake mechanism for sensor nodes to save energy. This approach finds the maximum distance node to calculate the maximum energy for data transmission. We considered it as a threshold energy $E_{th}$. Every node first checks its energy level for data transmission. If the energy level is less than $E_{th}$, it cannot transmit data. We also explain mathematically the energy consumption and average energy saving of sensor nodes in each round. Extensive simulations showed that when use this approach for WSNs significantly helps to extend the network lifetime and stability period.

## 1. INTRODUCTION

WSNs consist of a large number of sensor nodes deployed in a sensor field for object monitoring either inside the field or near the field. Recent advances in Microelectromechanical Systems (MEMS) based technologies have enabled the deployment of a large number of tiny sensor nodes with limited battery life time. These nodes have the low computational ability and small internal memory. These small sensor nodes capable of monitoring, sensing, aggregation and transmission of data to the sink. WSNs are used in many communication applications, including security, medical,surveillance and weather monitoring. Sensor nodes are able to measure various parameters of the environment and transmit collected data to the sink directly or through multi-hop communication. In WSNs sensor nodes cannot be replaced or recharged after deployment while most applications have pre-specified requirements.

Xue and GanzChen [1], define the lifetime when all node die. In [1, 2], it is defined as the time when the amount of dead nodes reaches a specified percentage (k% Die Time, KDT), when k=100 is all died time and half died time when k=50. So, the network lifetime is of great interest for the research in sensor networks [3, 4]. Nodes deployment is the first step in establishing the sensor network. Sensor nodes are battery powered and randomly deployed in target area. After the deployment sensor network cannot perform manually. Optimizing the energy consumption is one of the major tasks in WSNs is to prolong the network lifetime. To address these issues much work has been done in this area during the last few years. If the sensor nodes are deployed uniformly, the sensor nodes near the sink send their own data as well as the date collected by other nodes away from the sink in multi-hopWSNs. In this case, the sensor nodes near the sink consume more energy than the node away from sin and die more quickly. As a result, the network will disconnect when 90% of nodes are alive having sufficient energy left unused [5]. In this work, we investigate and try to remove the energy hole problem. We analyse the energy imbalance in these protocols and present sleep and awake mechanism to enhance the network lifetime for many to one WSNs. We use sleep and awake process to eliminate the energy hole problem. Further we conduct extensive simulations to investigate and confirm the performance of these techniques. Simulation results show the network lifetime and the total number of nodes in the network area use to enhance the network lifetime.

## 2. RELATED WORK

Various schemes have been proposed to address the Energy Hole Problem (EHP). Here we briefly describe the existing work related to our work, some of which are discussed in section I. One of the most fundamental issue is density control has attracted a great attention. In [5], the authors present a model for balanced density control to avoid energy holes. They use equivalent sensing radius and pixel based transmission schemes for balanced energy consumption. By activating different energy layered nodes in non-uniform distribution the energy holes problem is mitigated effectively. In [6], the authors purpose a Voronoi diagram-based distribution model for sensor deployment. In this model each node calculates its Voronoi polygon to detect coverage hole and move towards a better position for maximum coverage in the field. In [7], the authors discuss the Corona-based sensor network model for balanced energy depletion due to many to one communication in multi-hop sensor networks. They use mobile sensors to heal the coverage hole created due to large data relaying near the sink. In [8], the authors propose a multiple sink model to divide the network load near the sink to avoid the energy hole. The decision of a multiple sink is based on the amount of data loads in sensor networks.

The coverage problem is also a main reason for Energy Hole Problem (EHP). In [9], the authors discuss the distributed localization problem, Optimal Geographic Density Control (OGDC) [9], for full coverage as well as connectivity. They prove that if communication range is twice the sensing range then complete coverage implies connectivity. In [10], the authors investigate the theoretical analysis of different problems in routing protocols. They study the analysis of first died time, all died time and different parameters use to enhance the lifetime of the network. They also discuss the energy consumption of sensor nodes in different regions. Time and space analysis of network in different states like state without any node death, state with some node death, and when all node die also discuss. Tang and Xu [6], discuss how to optimize the network lifetime and data collection at the same time. Large amount of data is given to the sink by nearby nodes and less data from the nodes that are far away from the sink in the previous study. For this a rate allocation algorithm LexicographicalMax-Min (LMM) for data gathering is proposed to maximize the data gathering amount and maximize the network lifetime under balance data gathering condition. In [11], the authors investigate the non-uniform node distribution technique for balance energy consumption in sensor networks. Nodes near the sink consume more energy due to large data load, so the deployment of sensor nodes is in a geometric progression way from outer most Corona in inner Corona. In [12], the authors investigate uneven energy dissipation in sensor networks. They consider uniform node distribution in the network, and all nodes transmit the same amount of data packet to sink. In [13], the authors investigate different type of energies used by sensor nodes. They show that how initial energy, battery ratio and battery capacity of a sensor node affects the lifetime of the sensor network. Previous papers cover the topics of energy holes and different techniques to mitigate them. They mainly discuss the energy hole avoiding techniques. They do not give any technique to determine the exact location of energy hole and the time of occurrence. In E-HORM, we investigate the EHP and propose a technique to remove this problem.

## 3. ENERGY HOLE PROBLEM IN WSNS

There are many phenomena that affect the functionality; sensing and communication in WSNs. We discuss here the characteristics and the effects of energy holes. Unbalanced energy consumption is a major issue in WSNs when nodes are randomly deployed in sensor networks. Sensor nodes in the network behave as a data originator and data router [14]. Nodes near the sink have a greater load of data and consume more energy than nodes away from a sink. Therefore, nodes near the sink deplete more energy and die quickly than other nodes, leading what is called EHP around the sink. In this situation, no more data will be transmitted to the sink. So the network lifetime ends due to more depletion of energy near the sink. More sensor nodes due to dense deployment in any region may overlap and increase the hardware cost. However dense deployment is another reason for the creation of holes problem in WSNs.In all current routing schemes using optimal path intermediate nodes in the routing path deplete their energy more quickly, which expand the area of an energy hole.

Using proper nodes deployment techniques, EHP can be reduced and thus extends the network lifetime of WSNs. EHP can be reduced through different node deployment techniques, mobile sink node [11], non uniform deployment [12], or variable transmission ranges [15], [16], but these deployment strategies bring much more management cost.

Table 1: Cluster Based Protocols

| Protocols | Main features |
|---|---|
| LEACH | Equal cluster radius; each node has the same probability as the cluster head; the cluster head sends data to the sink directly. |
| TEEN | Equal cluster radius; each node has the same probability as the cluster head; the cluster head sends data to the sink directly. |
| DEEC | Equal cluster radius; multilevel heterogeneous network; advance nodes have more probabilities as the cluster head compares to normal modes; the cluster head sends data directly to the sink. |
| SEP | Equal cluster radius; two-level heterogeneous network; advance nodes have more probability as a cluster |

| | head as compare to normal nodes; cluster heads send data directly to the sink. |

## 4. ENERGY CONSUMPTION MODEL

We have found that due to EHP network dies early. In [5], the authors show that due to more depletion of energy near the sink nodes die more quickly, and network lifetime is over, even when up to 90% of energy is left unused. *Yifeng et al.* [17] deals with life time of sensor networks. They assume that the communication between nodes consume more energy as compared with data aggregation and data reception. Previous research shows that traffic near the sink is heavier than the traffic away from the sink. Energy consumption near the sink is greater and energy holes occur leading to the death of the entire network. This phenomenon reduces the lifetime of the whole network, because the energy holes are due to large energy consumption near the sink. If more nodes are deployed near the sink, there will be more nodes use to relay the distant data and hence extends the network lifetime which is also non-uniform node distribution strategy. The phenomenon of an energy hole makes the researchers realize that the network lifetime is determined by the weakest node while the region, size and time of energy holes are the temporal characteristics of the network lifetime. How to avoid EHP becomes an important research area now days. We use the same energy consumption model as used in [5], which is the first-order radio model. In this model radio dissipates $E_{elec}$ = 50nj/bit to run the transmitter circuitry, and $E_{amp}$ = 100pj/bit/m2 for the transmitter amplifier. The radio has the power to control and can adjust power according to the distance. We use free space $E_{fs}$ and multipath $E_{amp}$ loss model in our scheme. Receiving is also a high cost operation therefore; number of transmission and receiving should be minimal. To receive the 1-bit message the radio expends $E_{Rx}$ = $LE_{ele}$ energy. LEACH, TEEN, DEEC and SEP use the same constants ($k, E_{fs}, E_{amp}$) for calculating energy cost.

## 5. E-HORM: PROPOSED SCHEME

We consider the scenario where nodes are randomly deployed in a given region. Some nodes are selected to be active and rest are in sleep mode to maintain sensing, coverage and connectivity. We deploy n = 100 nodes randomly in a square area. The position of the sink is at the centre of the network. The energies of all the nodes are equal while the energy of the sink is unlimited. In our model, nodes transmit data to the sink based on residual energy and the distance between nodes and sink. E-HORM scheme has three major phases: initializing phase, threshold calculating phase, cluster formation and sleep/awake scheduling phase. In every round sink first checks the maximum distance node in the field. It then calculates the required energy to transmit data to the sink. We set this energy as a threshold energy $E_{th}$. In every round if the energy level of a node is greater than or equal to threshold energy, the node transmits data to the sink.

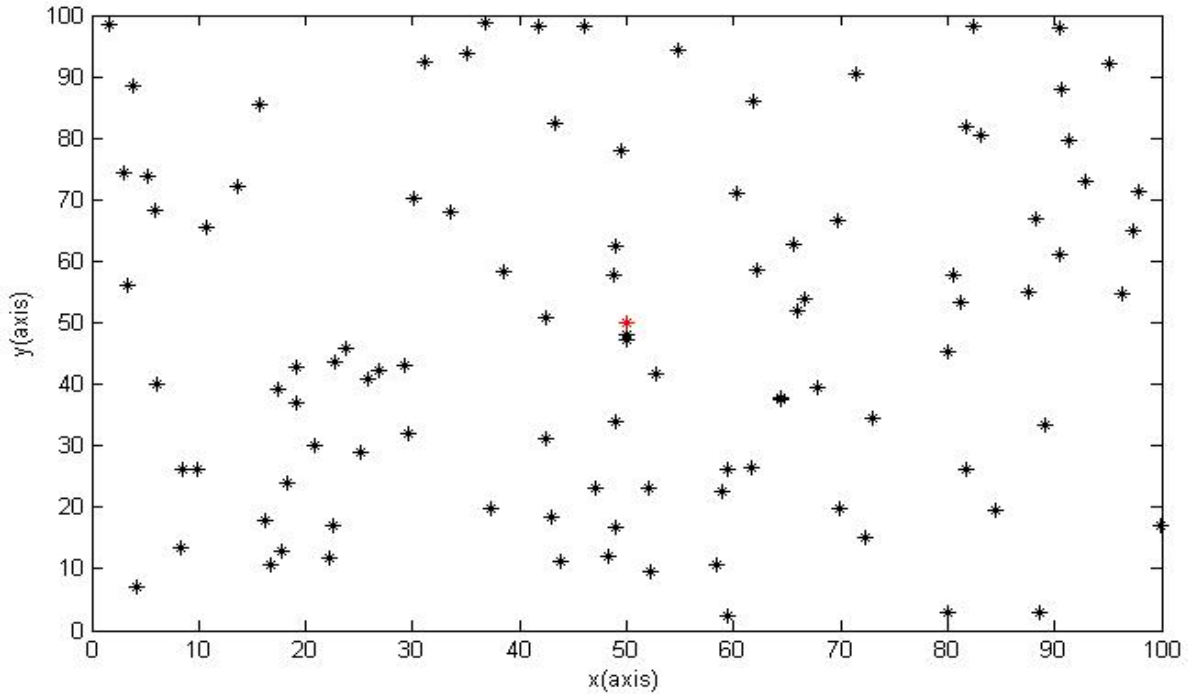

Fig. 1. 100-node random network

If the energy level of any node is less than the threshold value, it cannot be transit data to the sink. When the energy level is less than the threshold energy value, it moves towards the sleep mode to save energy. In the next round again sink calculates the energy level of each node. The same process repeats until the number of sleep nodes are equal to NS = 10. Now we discuss the idea of awake mechanism. When numbers of sleep nodes are greater NS> 10 than the node which is the first one in the sleep position move towards awake position. In next round when the number of nodes in sleep position is greater NS> 11 than the node which is in second position of the sleeper position moves towards the active position and so on. When the number of sleep modes NS < 10 then the node whose energy level is less then threshold energy level moves towards sleep position. In this scenario, the total node in the sleep position is again equal to NS = 10. Flowchart describes the sleep and awake mechanism. We implement our scheme in the cluster based protocols L ACH, TEEN, DEEC and SEP. Nodes choose their cluster head on the basis of predefined probability. The cluster heads broadcast their status that each node can determine to which cluster head it wants to associate to consume minimum energy for data transmission. After association, each cluster head creates a schedule to the nodes in its cluster. Sink assigns TDMA slots to every node. Nodes only transmit their data during their assigned TDMA slots. This is to save energy for the nodes that are in sleep mode except during transmission.

*A. Sensor Node Sleep Scheduling*

Before performing the sleep schedule, we examine the energy level of each node according to their distance from the sink and the following steps.

- Case 1: E0 > Eth: When remaining energy is greater than the threshold energy, the node is in active mode and ready for communication.

- Case 2: E0 < Eth: When remaining energy is less than the threshold energy, the node moves towards sleep position.

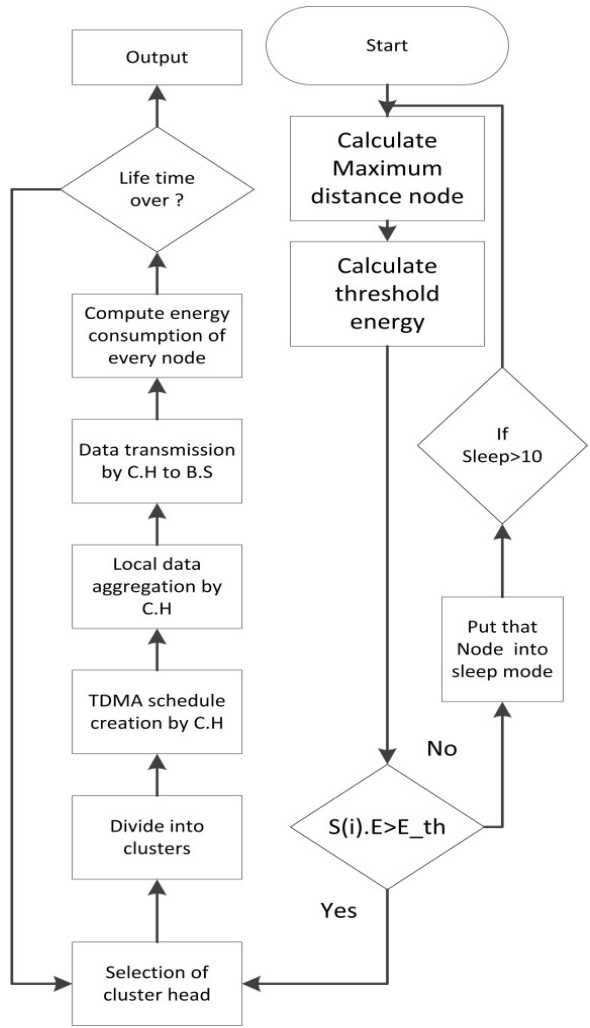

Fig. 2. E-HORM Flowchart

In Fig. 2 we calculate the value of threshold energy of each node for sleep and awake mechanism. Sink selects the node to put into sleep position randomly. Each node sets the sleeping scheduling according to the threshold energy.

To calculate the threshold energy, we use the following formula,

$$E_{th} = ((ETX + EDA) * D) + (E_{amp} * D * d^4) \qquad (1)$$

Where $E_{th}$ is threshold energy, D is the length of the data packet and d is the distance between maximum distance node and sink. Sink calculates the threshold energy for sleep and awake mechanism. $CH_2$ is the more distant node and consumes more energy for transmission. We calculate the threshold energy of this node according to the given formula.

*B. E-HORM Formulation*

Based on the network model, the nodes belonging to CH forward both the data generated by themselves and the data generated by its member nodes. Nodes which are not CH need not forward any data. Suppose nodes are randomly deployed in the network and there is no need for data aggregation at any forwarding node. Based on transmission mechanism data for CH receive and forward is $(D_1+D_2+D_3+....+D_N)$ and $(D_{CH}+D_1+D_2+D_3+....+D_N)$.

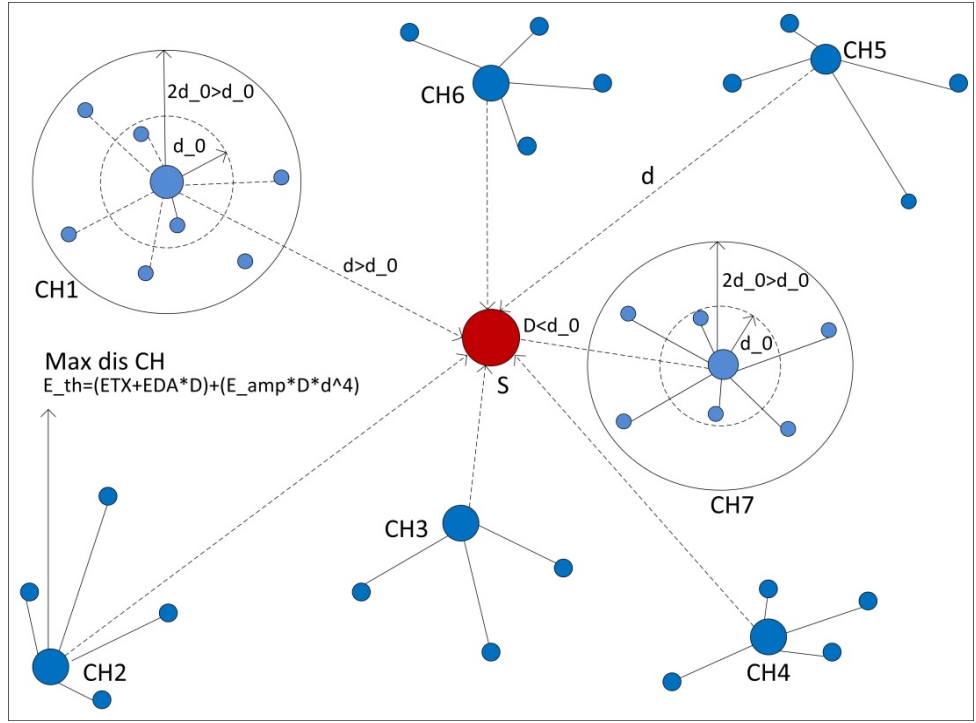

Fig. 3. Threshold energy of distant node

In the above figure if the distance between N and CH is d<d_0 than energy consumption for data transmission from N to the CH.

$$E_N^{CH} = D_N^{CH}(E_{ele}) + D_N^{CH}(E_{fs})(d^2) \quad (2)$$

Where d is the distance between N and CH. The nodes send its own data to CH for each data-gathering process in different rounds near the sink. Now considering the scenario where the distance between N to CH is d>d_0.

$$E_N^{CH} = D_N^{CH}(E_{ele}) + D_N^{CH}(E_{amp})(d^4) \quad (3)$$

Energy consumed by CH to transmit data to the S when distance between them is d<d_0

$$E_{CH}^S = D_{CH}^S(E_{ele}) + E_{DA} + D_{CH}^S(E_{fs})(d^2) \quad (4)$$

When the distance between CH and S is d>d_0

$$E_{CH}^S = D_{CH}^S(E_{ele}) + E_{DA}D_CH^S(E_{amp})(d^4) \quad (5)$$

$$E_{Total\_CH} = E_{CH} + E_N \quad (6)$$

$$E_{Average\_CH} = \frac{E_{Total\_CH}}{N} \quad (7)$$

Energy saving in each round for the normal node

$$E_{Save\_N} = E_{elec} + E_{TX} + E_{amp} \quad (8)$$

Energy saving for CH is

$$E_{Save\_CH} = E_{ele} + E_{DA} + E_{TX} + E_{RX} + E_{amp} \quad (9)$$

Energy saving for all sleep nodes

$$E_{Save\_Total} = \sum_{i=0}^{n} E_i \quad (10)$$

Where n is the total number of nodes that are in sleep mode

Average energy saving is

$$E_{Save\_Average} = \frac{E_{Save\_Total}}{N} \quad (11)$$

Fig. 3 shows CHs with different distances from the sink. Each cluster head contains different numbers of sensor nodes. $CH_1$ shows two regions. Distance between CH and N is less than $d_0$ in the dotted region. The energy required to transmit data from N to CH is less than the energy required for the node that is outside of the dotted region. This is because the distance between N outside the dotted region is greater than $d_0$. The energy required to transmit data from $CH_1$ to S is greater because the distance is $d > d_0$. In $CH_7$ distance between the S and CH is less than $d < d_0$. Energy used to transmit data from $CH_7$ to S is less than CH1.

*C. Analysis*

We calculate the sleep probability of sensor nodes for each round by using threshold energy $E_{th}$ of distant nodes. The nodes away from the sink increase the sleep probability. In this way, the nodes consume approximately balance energy to enhance the network lifetime.

For WSNs the sleeping scheduling is very important due to limited energy of sensor nodes. If a node set into the active position for a long time, it consumes a lot of energy. On the other way, the transmissions create more delay for long time sleep duration. In this paper, we design an optimum sleeping control mechanism to avoid both of the problems.

6. SIMULATION RESULTS

Our results are based on analyses and are validated by Matlab simulations. In our simulations, we consider a sensor network with 'n' number of homogeneous or heterogeneous sensor nodes, which are randomly deployed in a square field. The only sink is located at the centre of the field. Sensor nodes do not move after deployment. Sensor nodes are a limited initial energy $E_{init}$, while the energy of the sink is unlimited. The transmission ranges of sensor nodes are adjustable according to the distance from the sink. All nodes need to send the data packets to the sink in a cycle time. In each round sensor nodes are selected to work and the rests of nodes are set to sleep mode to save energy. In E-HORM, this mechanism is called sleep awake process. We apply our scheme in four categories of the cluster based protocols LEACH, TEEN, DEEC and SEP. In this paper the cluster based network refers to the routing protocols.

All parameter values use in our simulations are listed below.

Table 2: Simulation Parameters

| Symbol | Description | Value |
| --- | --- | --- |
| $X_m$ | Distance at x-axes | 100 meters |
| $Y_m$ | Distance at y-axes | 100 meters |
| N | Total number of nodes | 100 Nodes |
| $E_0$ | Total energy of network | 0.5 j |
| P | Probability of cluster head | 0.1 |
| $E_{RX}$ | Energy dissipation: receiving | 0.0013/pj/bit/m^4 |
| $E_{fs}$ | Energy dissipation: free space model | 10/pj/bit/m^2 |
| $E_{amp}$ | Energy dissipation: power amplifier | 100/pj/bit/m^2 |
| $E_{ele}$ | Energy dissipation: electronics | 50nj/bit |
| $E_{TX}$ | Energy dissipation: transmission | 50/nj/bit |
| $E_{DA}$ | Energy dissipation: aggregation | 5/nj/bit |
| $d_0$ | Reference distance | 87 meters |
| n | Number of sleep nodes | 10 Nodes |

In this section, we evaluate and explain the experimental results of our purposed work. Two main matrices, including the stability period and network survival lifetime are measured and compared with the existing work. We write a Matlab program to simulate our work with initial deployment of nodes. We explore the theoretical analysis of node distribution according to random deployment. We find that it is difficult to remove all energy holes due to random deployment nature of sensor nodes in WSNs. We describe that with the purposed transmission scheme, better results are achieved.

We evaluate and compare our purposed technique with LEACH, TEEN, DEEC and SEP protocol with random node deployment of nodes in the sensor network. Our technique outperforms in terms of network life time and stability period. Now we explain how the sleep and awake mechanism are carried out in LEACH to remove the energy hole in WSNs.

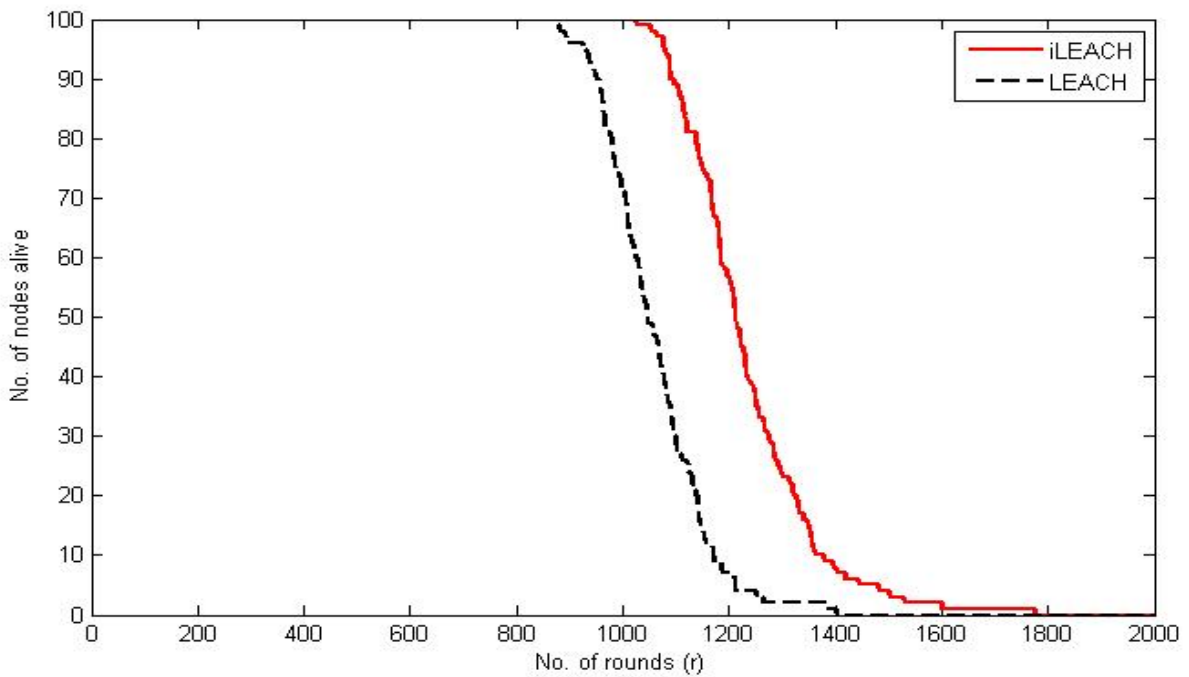

Fig. 4. Number of alive nodes in different target areas

LEACH is homogeneous routing protocol. In LEACH all nodes have the same probability to become a cluster head. Cluster heads consume more energy than normal nodes. In every round, all nodes participate in the communication process. Node that is far away from the sink and also a cluster head consume more energy than the nodes that are near the sink node. The main reason for more energy consumption is that the cluster head sends its own as well as aggregated data collected from its members. Node forward its own data to cluster heads according to TDMA schedule. Sink assigns slots to every node for data transmission. All nodes are in sleep node and turn on during transmission to save energy.

According to our approach, the nodes that have the energy level less than the threshold are in the sleep mode to save energy. In this way, we save energy to prolong the network life time and stability period. The simulation results of iLEACH and LEACH is shown in the above figures. The life time and stability period of the network are greater in iLEACH.

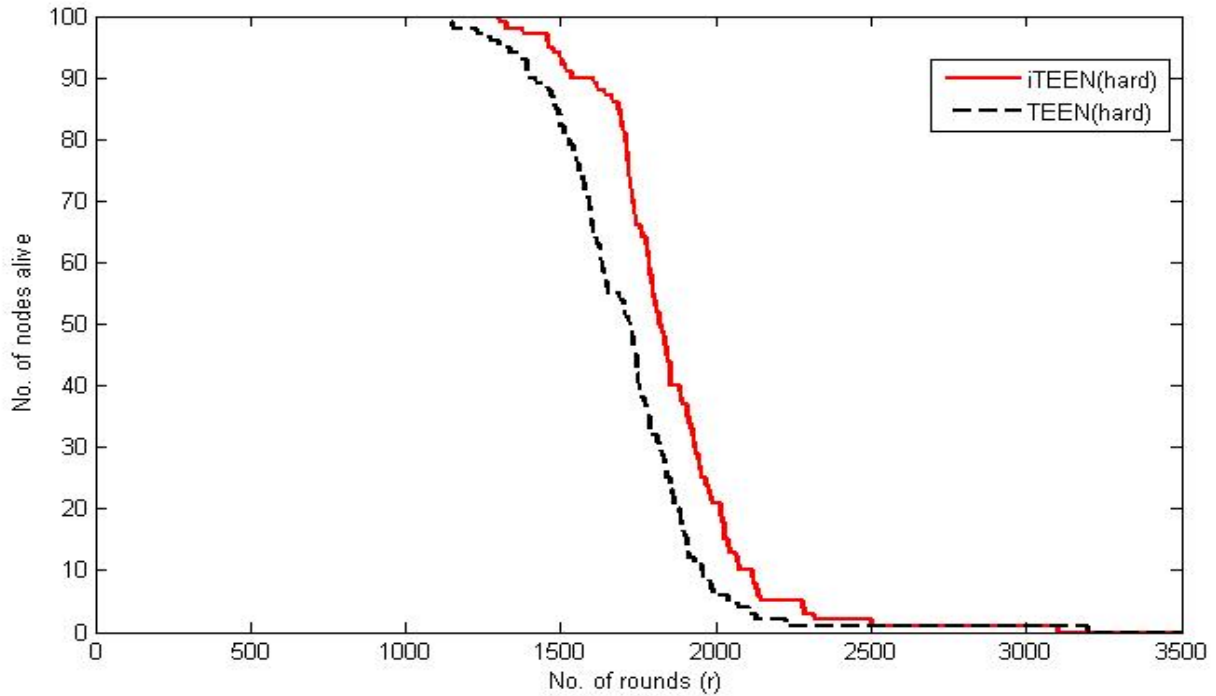

Fig. 5. Number of alive nodes in different target areas

SEP is a heterogeneous clustering protocol. In this section, we compare the performance of SEP and our proposed protocol iSEP in the same heterogeneous setting. Extra energy is distributed over-all advance nodes in the field. This setting latter provides the more stable region and network lifetime. Fig. 5 shows the result of SEP and our proposed scheme iSEP. It is very clear that the stable region of iSEP is greater than SEP even though the network lifetime is not very large. Due to balanced energy consumption network, stability increased then SEP. In SEP when the first node dies the system becomes unstable due to population reduction. The death ratio of normal nodes is greater than advance nodes. This is because normal nodes have less energy than advance nodes. In sleep and awake mechanism, the node which is far away from the base station and consumes more energy put into the sleep node to avoid an energy holeproblem. In each round, the nodes having less energy than threshold reference energy put into sleep mode to save energy. This mechanism enhances the stability period of SEP due to better utilization of energy.

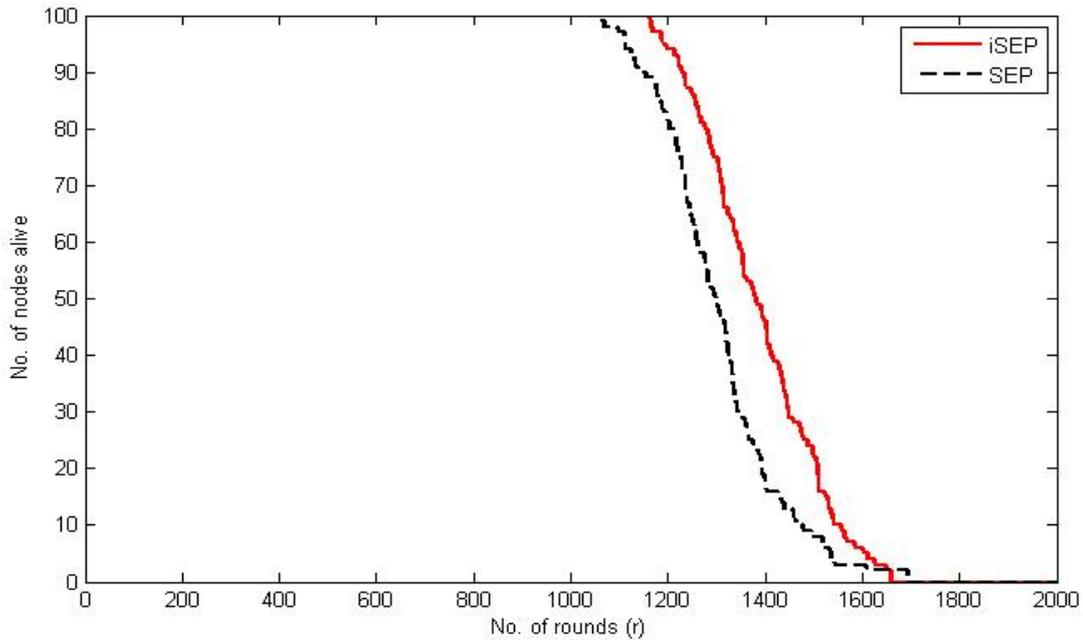

Fig. 6. Number of alive nodes in different target areas

Here we evaluate and compare the performance, of DEEC and iDEEC protocols. For the best evaluation of performance, we ignore the disturbances due to signal collision and interference in a wireless medium. It is obvious that the network lifetime of our proposed scheme is greater compared to simple DEEC protocols. There is a minute change in the stability period of SEEC and iDEEC. This is because the energy of each node is different from other nodes. In periodically sleep and awake mechanism, we can best utilize the energy consumption in each region to remove energy holes.

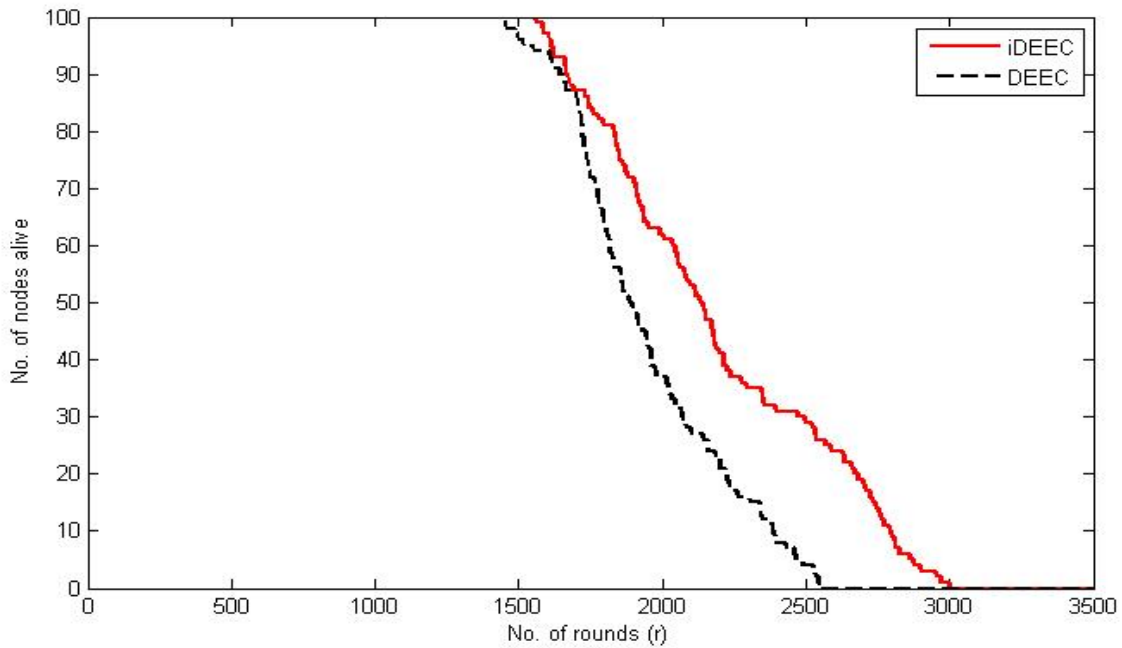

Fig. 7. Number of alive nodes in different target areas

### 7. CONCLUSION

In this article, we focus on the energy hole problem and energy consumption in these protocols. We discussed the creation of energy holes in homogeneous and heterogeneous routing protocols. We implement our approach in LEACH, TEEN, DEEC and SEP routing protocol. Due to random deployment in these protocols, there exists the probability of energy holes. Sleep and awake mechanism to remove energy holes in WSNs is proposed. We investigated that after our proposed scheme, a better energy consumption is achieved. As for network lifetime, this work clearly gives the results in terms of network lifetime and stability period. Sensor nodes consume balance energy, and hence maximize the network lifetime. This paper clearly points out how we can remove the energy hole problem in WSNs, and other researchers can also easily propose a new protocol according to deployment techniques, which avoid the energy holes. Simulation results show that the results of my scheme perform better than previous schemes in terms of network life time and stability period.